# System Characterization of Dispersive Readout in Superconducting Qubits

Daniel Sank, Alex Opremcak, Andreas Bengtsson, Mostafa Khezri, Zijun Chen, Ofer Naaman, and Alexander Korotkov
*Google Quantum AI, Goleta, CA 93111, USA*

Designing quantum systems with the measurement speed and accuracy needed for quantum error correction using superconducting qubits requires iterative design and test informed by accurate models and characterization tools. We introduce a single protocol, with few prerequisite calibrations, which measures the dispersive shift, resonator linewidth, and drive power used in the dispersive readout of superconducting qubits. We find that the resonator linewidth is poorly controlled with a factor of 2 between the maximum and minimum measured values, and is likely to require focused attention in future quantum error correction experiments. We also introduce a protocol for measuring the readout system efficiency using the same power levels as are used in typical qubit readout, and without the need to measure the qubit coherence. We routinely run these protocols on chips with tens of qubits, driven by automation software with little human interaction. Using the extracted system parameters, we find that a model based on those parameters predicts the readout signal to noise ratio to within 10% over a device with 54 qubits.

## I. INTRODUCTION

Quantum computing systems are increasing in scale and accuracy. Error correction protocols sit on top of an underlying physical level comprised of individual qubits, each of which is an analog device with many physical parameters. Can those parameters be extracted reliably, and can the full system performance be predicted from a model based in measurements of those parameters? A demonstration of beyond classical computation in superconducting qubits [1] answered this question affirmatively, showing that the cross entropy in a random circuit sampling experiment was well predicted by a model based on single- and two-qubit component metrics. More recently, the same was shown in experiments with a primordial surface codes where logical error rates were well predicted by model based component metrics [2, 3]. Such precise agreement between prediction and experiment relies on accurate characterization. Additionally, an accurate model of the readout system is an essential ingredient in tuning up the quantum processor for best performance [4].

Of course, the value of accurate characterization extends beyond predicting the performance of and tuning up existing devices. Fault tolerant quantum computation requires faster, more accurate, and lower leakage readout than has been reported to date in a system with sufficient multiplexing. Achieving the required speed-up, while isolating the qubits from decay through the readout system (via e.g. the Purcell effect [5–8]), spectral crosstalk, and dephasing imparted by noise in the readout resonators [9–13], requires improved circuits. As we set ourselves to designing those new circuits, we want to know that the models in which we root our design are founded in agreement with experiment.

To this end, we have developed a set of experiments which extract the parameters of the dispersive readout system in a superconducting qubit device. To satisfy the needs of increasingly larger devices, these experiments have been designed with automation in mind; they require as few as possible prerequisite calibrations, run in parallel and in-situ, and require only simple model fitting. We regularly run these experiments on chips with tens of qubits via automation software with little human interaction.

The paper is organized as follows. Section II reviews a simple and well known model of dispersive measurement. Section III describes a single experiment which, using the qubit as a detector of the resonator energy, extracts the resonator dispersive shift, resonator linewidth, and resonator drive power. Section IV presents two additional ways to measure the resonator linewidths and shows an example wherein increased and reliable measurement statistics revealed the root cause of a systematic design error. Section V introduces a protocol to measure the readout efficiency without measuring the qubit coherence. Section VI compares measured readout performance with performance predicted from a model based on our component metrics, finding ±10% accuracy in signal to noise ratio (SNR). Finally, section VII offers outlook and conclusions. The experiments described here are part of the readout optimization procedure described in detail in Ref. [4] and used in Refs. [1] and [2].

## II. MODEL

We study the system illustrated in Fig. 1 (a) in which several qubit-resonator pairs are coupled to a common bandpass Purcell filter. As discussed in Ref. [7], the dynamics of each resonator can be accurately described by the simplified circuit shown in Fig. 1 (b) with the resonator directly coupled to the input/output line. In this picture, the effect of the Purcell filter is to rescale the linewidth $\kappa$ of the resonator. In a frame rotating with frequency $\omega_{\text{frame}}$, and in the rotating wave approximation, a single driven resonator is described by the pair of

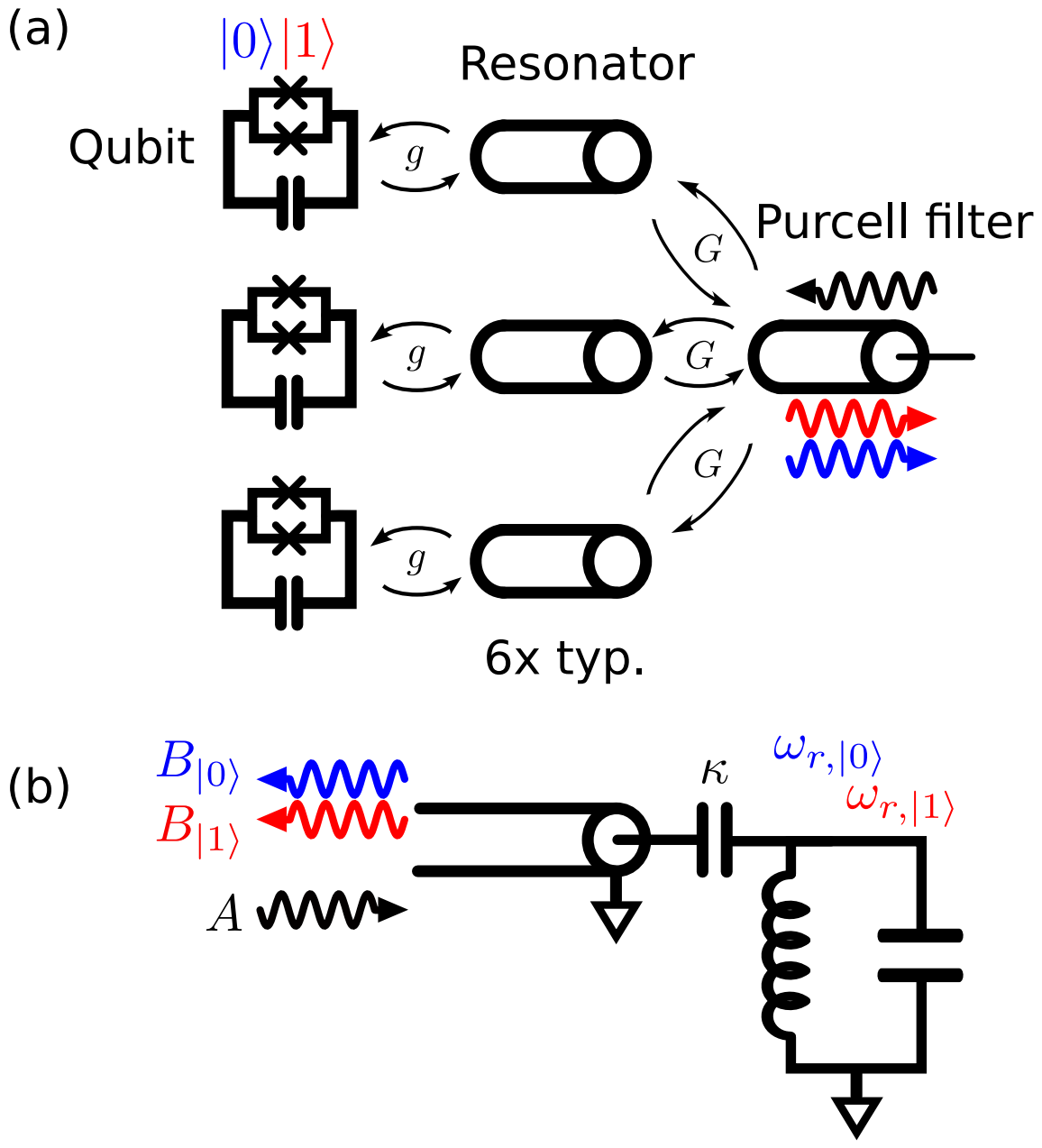

FIG. 1. Schematic diagram of the readout system. (a) The readout circuit includes a shared Purcell filter coupled to six readout resonators. Each readout resonator is coupled to a transmon qubit. The $|0\rangle$ (blue) and $|1\rangle$ (red) states of the transmon correspond to different inductances of the SQUID, which lead to different phases of the reflected signals. (b) The readout circuit can be approximated as a single resonator, with an effective linewidth $\kappa$, whose frequency depends on the qubit state.

equations

$$\begin{aligned}\dot{\alpha}(t) &= -i\Delta\alpha(t) - \frac{\kappa}{2}\alpha(t) + \sqrt{\kappa}A(t) \\ B(t) &= -A(t) + \sqrt{\kappa}\alpha(t)\end{aligned} \tag{1}$$

where $\alpha$ is the complex resonator amplitude, $\omega_r$ is the resonator frequency, $\Delta \equiv \omega_r - \omega_{\text{frame}}$, is the detuning of the resonator from the frame, $\kappa$ is the resonator's linewidth (assumed to be set by the coupling to the drive line (in this work $1/\kappa \approx 30\,\text{ns}$), and $A(t)$ and $B(t)$ are the complex amplitudes of the incident (drive) and reflected (measured) waves at the point that the input-output transmission line couples to the resonator. $A$ and $B$ are both in the rotating frame. See Fig. 1 (b). In these equations, $A$ and $B$ are normalized such that $|A|^2$ and $|B|^2$ are photon flux (dimensions of inverse time), and $|\alpha|^2 = \bar{n}$ is the mean number of photons in the resonator, which is assumed to always be in a coherent state [14]. The resonator frequency depends on the state of the qubit; the qubit states $|0\rangle$ and $|1\rangle$ shift the resonator to frequencies $\omega_{r,|0\rangle}$ and $\omega_{r,|1\rangle}$ with difference $\omega_{r,|1\rangle} - \omega_{r,|0\rangle} = 2\chi < 0$ and "middle frequency" $\omega_{\text{r,m}} \equiv (\omega_{r,|0\rangle} + \omega_{r,|1\rangle})/2$.

The different values of resonator frequency for the two qubit states lead to different outgoing waves $B$, enabling qubit state measurement. For each measurement shot, we amplify and digitize the outgoing wave and compute a complex number

$$Z = C\int_0^T B(t)W(t)\,dt \tag{2}$$

where $W(t)$ is a dimensionless weight function chosen to maximize the signal to noise ratio (SNR) [15, 16], $T$ is the total integration time, and $C$ is a constant capturing the conversion from the physical signal to digitized samples. The two qubit states $|0\rangle$ and $|1\rangle$ correspond to two different outgoing waves $B_{|i\rangle}$ with two associated points $Z_{|i\rangle}$. The distance in the complex plane between $Z_{|0\rangle}$ and $Z_{|1\rangle}$ is

$$\begin{aligned}\left|Z_{|0\rangle} - Z_{|1\rangle}\right| &= C\left|\int_0^T (B_{|0\rangle}(t) - B_{|1\rangle}(t))W(t)\,dt\right| \\ &= C\sqrt{\kappa}\left|\int_0^T (\alpha_{|0\rangle}(t) - \alpha_{|1\rangle}(t))W(t)\,dt\right| .\end{aligned} \tag{3}$$

Noise in the system adds random displacements to $Z$. Typically, these displacements are described by a 2-dimensional Gaussian distribution. As discussed in detail in Appendix A, the 1-dimensional width $\sigma$ of this distribution is effectively given by $2\sigma^2 = C^2(1/2)\int_0^T |W(t)|^2\,dt$. The SNR is therefore

$$\begin{aligned}\text{SNR} &\equiv \frac{\left|Z_{|0\rangle} - Z_{|1\rangle}\right|^2}{2\sigma^2} \\ &= \eta\kappa\frac{\left|\int_0^T \left(\alpha_{|0\rangle}(t) - \alpha_{|1\rangle}(t)\right)W(t)\,dt\right|^2}{(1/2)\int_0^T |W(t)|^2\,dt}\end{aligned} \tag{4}$$

where the quantum efficiency $\eta < 1$ describes the reduction in SNR due to loss and added noise as the signal travels from the readout resonator to the detector [17–19]; $\eta = 1$ is possible only with a phase sensitive amplifier and is limited to $\eta \leq 0.5$ with a phase preserving amplifier as is used in this experiment. The remainder of this paper describes measurements of the parameters appearing in Eq. (1) and comparison of Eq. (4) against experiment.

For some measurements, it is convenient to operate in the steady state, where the drive $A$ is monochromatic, has constant magnitude, and frequency equal to $\omega_{\text{frame}}$, so that $\dot{\alpha} = 0$. Solving Eq. (1) in that case, we find

$$\alpha = \frac{2A}{\sqrt{\kappa}}\frac{1}{1 + i2\Delta/\kappa} \tag{5}$$

with photon occupation

$$\bar{n} = |\alpha|^2 = \frac{\kappa\,|A|^2}{\Delta^2 + (\kappa/2)^2} . \tag{6}$$

## III. DISPERSIVE SHIFT, LINEWIDTH, AND POWER

Equation (4) for the SNR and related equations for the outgoing field $B$ have four parameters: the dispersive shift $\chi$, the resonator linewidth $\kappa$, the drive strength $A$, and the efficiency $\eta$. We measure the first three of these parameters in a single experiment based on the ac Stark effect, using the pulse sequence shown in Fig. 2 (a). The qubit begins at the idle frequency $\omega_{\mathrm{q},0}$. We prepare the qubit in either $|0\rangle$ or $|1\rangle$ and then drive the resonator at a frequency $\omega_{d,\mathrm{r}}$ ("omega drive resonator") for a duration much longer than $1/\kappa$ so that the resonator is driven into the steady state with a certain photon occupation $\bar{n}$. This resonator occupation causes the qubit frequency to shift (ac Stark shift) by $\delta\omega_\mathrm{q} \approx 2\chi\bar{n}$ [20]. We then apply a pulse to the qubit at a frequency $\omega_{d,\mathrm{q}}$ ("omega drive qubit") which flips the qubit state if $\omega_{d,\mathrm{q}} = \omega^*_{d,\mathrm{q}} \approx \omega_{\mathrm{q},0} + \delta\omega_\mathrm{q} = \omega_{\mathrm{q},0} + 2\chi\bar{n}$. Finally, we allow the resonator to ring down, and then measure the state of the qubit using dispersive readout through that same resonator. The frequencies $\omega_{d,\mathrm{r}}$ and $\omega_{d,\mathrm{q}}$ are swept over a two-dimensional grid and for each grid point we record the probability that the final qubit state was different from the one prepared at the start of the pulse sequence. In other words, we record the probability that the qubit probe pulse frequency matched the ac-Stark-shifted frequency of the qubit in the presence of the resonator drive. Therefore, we are using the ac Stark shifted qubit frequency as a sensor of the resonator's internal energy. This protocol requires $1/\kappa \ll T_1$ where $T_1$ is the lifetime of the qubit state $|1\rangle$.

Results are shown in Fig. 2 (b). Typical values of $\omega_{d,\mathrm{r}}/(2\pi)$ are $\omega_{\mathrm{r,m}}/(2\pi) \pm 10\,\mathrm{MHz}$ and typical values of $\omega_{d,\mathrm{q}}/(2\pi)$ are $\omega_{\mathrm{q},0}/(2\pi)^{+10\,\mathrm{MHz}}_{-30\,\mathrm{MHz}}$. Taking our value for $\bar{n}$ from Eq. (6) with $\Delta = (\omega_{\mathrm{r,m}} \mp \chi) - \omega_{d,\mathrm{r}}$, we find the qubit drive frequency where we expect to see the qubit change state,

$$\omega^*_{d,\mathrm{q}} = \omega_{\mathrm{q},0} + 2\chi \frac{\kappa\,|A|^2}{(\omega_{d,\mathrm{r}} - (\omega_{\mathrm{r,m}} \mp \chi))^2 + (\kappa/2)^2} \tag{7}$$

where the upper and lower signs are for the qubit in $|0\rangle$ and $|1\rangle$. Equation (7) describes a pair of Lorentzian curves with means $\omega_{\mathrm{r,m}} \mp \chi$, depths $(2\chi)(4\,|A|^2/\kappa)$, and widths $\kappa$. We fit both curves, finding $2\chi$ as the horizontal distance between their means and $\kappa$ as their widths. Then with $\chi$ and $\kappa$ known we find $|A|^2$ from their depths. We assume that $|A|^2$ is proportional to the pulse generator power [21], so this calibration can be understood as a calibration between generator power and power incident on the resonator. The Lorentzian fits are shown overlaid with the raw data in Fig. 2 (b), and the two fits are shown together in Fig. 2 (c).

This experiment which we call "chi-kappa-power" (CKP) is easy to calibrate because it requires only crude knowledge of the resonator and qubit frequencies, crude calibration of the qubit control pulses and readout, and requires no calibration of microwave scattering parameters or changes to the apparatus. Notably, the two independent variables are both frequencies requiring no calibration, and we have no strong requirement on the contrast in the dependent variable (the qubit excitation probability). The traditional method of fitting the phase and magnitude of a wave scattered from the readout system suffers the complexity that, in a multiplexed system with a shared Purcell filter, the frequency proximity of neighboring resonators requires fitting the entire system to a model with many parameters. On the other hand, because each resonator is coupled to exactly one qubit, the qubit ac Stark shifts measured in CKP show the simple Lorentzian profile seen in Fig. 2. Simulation of the CKP experiment shows that it extracts the correct values of $\chi$, $\kappa$, and $A$ to within a few percent.

The CKP experiment does have some basic prerequisites. Before running it, the qubit pi pulses and basic readout are calibrated as described in Ref. [22], from which we learn the upper limit of $\omega_{d,\mathrm{q}}$ and the center value of $\omega_{d,\mathrm{r}}$. The qubit pulse duration is chosen to be 100 ns and its amplitude is scaled so that the pulse area is equal to 0.8 of the area in a pi pulse. To choose the resonator drive power $A$ in the CKP experiment, we drive the resonator at $\omega_{r,|0\rangle}$ and find the power at which the observed qubit ac Stark shift is $-20\,\mathrm{MHz}$.

## IV. OTHER MEASUREMENTS OF KAPPA

The CKP experiment measures $\kappa$ in the frequency domain as resonator line widths, but $\kappa$ is also the energy ringdown rate of the resonator in the time domain. We measure $\kappa$ in the time domain with two different protocols: 1) observing the time dependent ringdown of the qubit ac Stark shift, and 2) observing the time dependent voltage leaked from the resonator itself.

### A. ac Stark ringdown

In this method, we use the qubit's ac Stark shift as a probe of the resonator energy, just as in the CKP experiment, but now in the time domain. The pulse sequence is similar to the CKP sequence. The qubit is left in state $|0\rangle$ and the resonator is driven for a duration several times $1/\kappa$ to put it into the steady state. The resonator drive is turned off, and then a pulse of variable frequency is applied to the qubit at a variable time. In this way, we find the qubit frequency as a function of time after the end of resonator drive. Because the energy in the resonator decays as $\exp(-\kappa t)$, and because the qubit ac Stark shift is approximately linear in the resonator energy, we see the qubit frequency relax exponentially to its idle value. Fitting this exponential curve gives us $\kappa$. Figure 3 (a) shows an example of the measured time dependent qubit ac Stark shift with exponential fit. We call this method "ac Stark ringdown". ac Stark ringdown has similar advan-tages as CKP: it avoids fitting an S parameter model to the multiplexed readout system and requires only crude knowledge of the qubit and resonator frequencies.

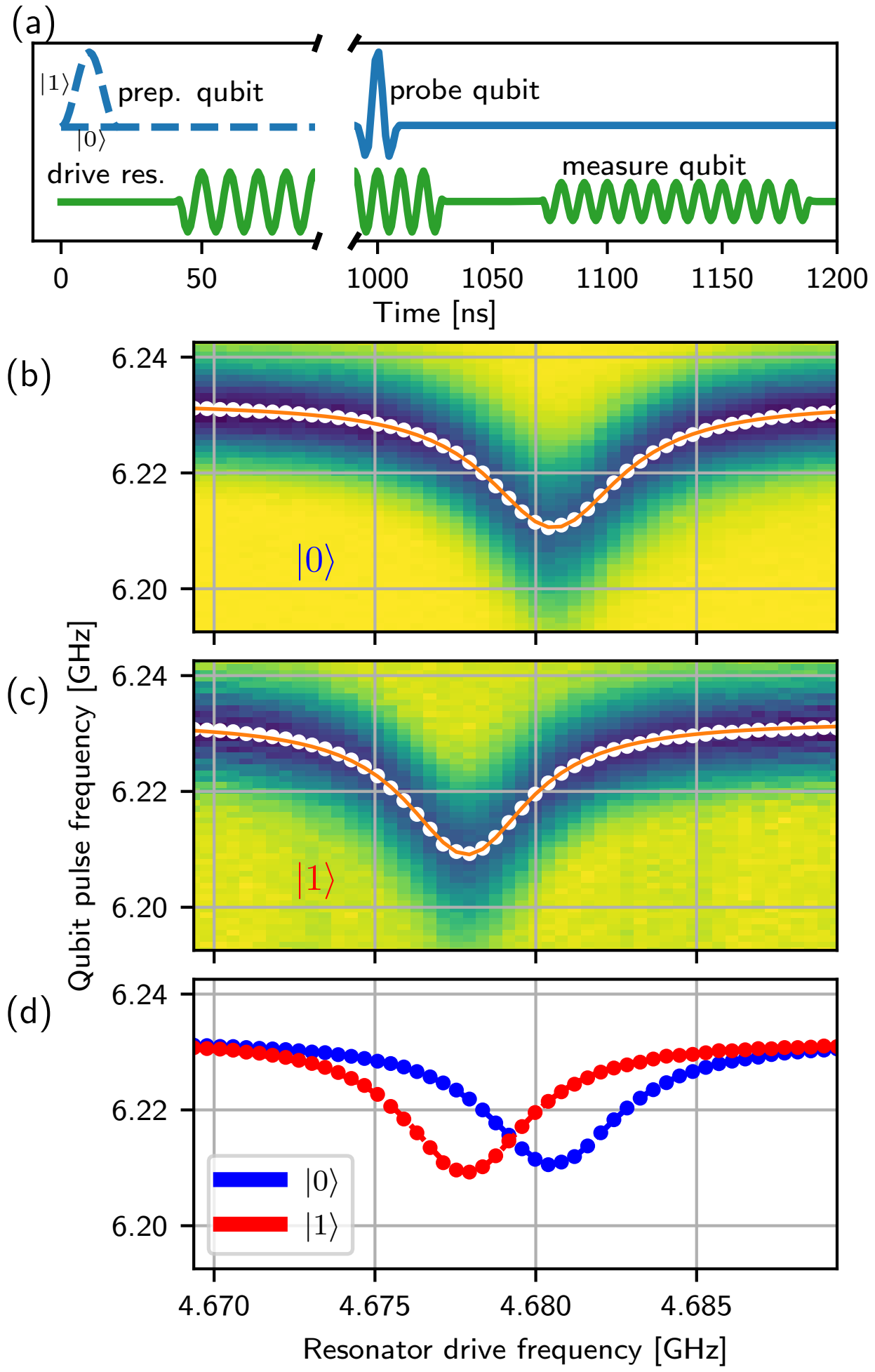

FIG. 2. The chi-kappa-power experiment. (a) Pulse sequence. Blue and green curves indicate qubit and resonator drives. The dashed line indicates that the qubit preparation pulse is absent(present) for preparation of $|0\rangle(|1\rangle)$. The first resonator pulse at a variable frequency populates the resonator and ac-Stark shifts the qubit, while the second resonator pulse at a fixed frequency reads out the final qubit state. (b) Probability of measuring the qubit in the opposite state from which it was prepared. The dark color bands, forming Lorentzian curves, indicate high probability. The white dots indicate the fitted minimum for each vertical slice of data, and the orange curves indicate the Lorentzian fits. (c) Lorentzian fits for $|0\rangle$ and $|1\rangle$.

### B. Voltage ringdown

The voltage ringdown method is similar to ac Stark ringdown, except that we measure the time dependent voltage arriving at the receiver and fit its exponential decay. The voltage decays with rate $\kappa/2$ instead of the energy decay rate $\kappa$. Figures 3 (b) and (c) show examples of the oscillating and decaying voltages collected by the I and Q channels of the receiver.

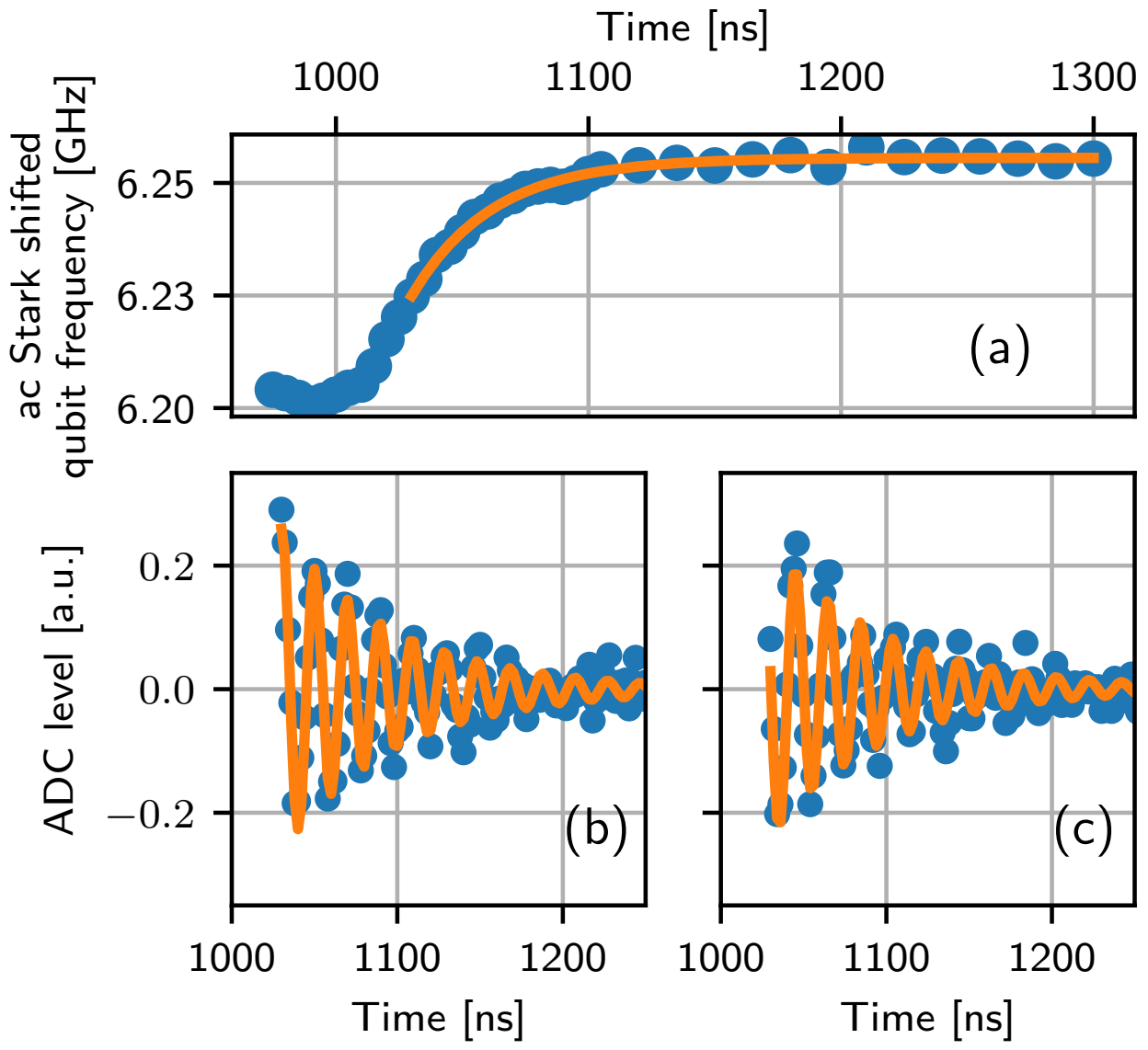

FIG. 3. Kappa measured in the time domain. (a) Time dependent ac Stark shift of the qubit. (b,c) Decaying oscillations of the voltage arriving at the receiver I (b) and Q (c) channel ADC's.

### C. Correlation between measurement methods

Figure 4 shows the correlations between our three methods for measuring $\kappa$. Evidently, the two methods based in the ac Stark effect agree the best, while the method using the ringdown of the received voltage agrees the least well with the other two.

### D. Distribution

The designed value of $\kappa$ for the device used in all measurements described so far was $(30\,\text{ns})^{-1}$, but as we can see in Fig. 4 the measured values range from $(40\,\text{ns})^{-1}$ to $(20\,\text{ns})^{-1}$, a factor of 2 distribution from smallest to largest. Such a large distribution poses a problem for device design. Any given requirement on readout duration implies a lower bound on $\kappa$. Suppose we designed such that our smallest realized $\kappa$ were at the design value of $1/(30\,\text{ns})$; then with a $2\times$ variation, our largest $\kappa$ would be $1/(15\,\text{ns})$ which would incur too much damping on the qubits [5, 6]. On the other hand, if we design defensively so that the fastest $\kappa$ were $1/(30\,\text{ns})$, then the slowest would be $1/(60\,\text{ns})$, meaning our readout system would be $\sim 2\times$ slower than intended. Because the readout is multiplexed, the slowest readout determines the repetition rate of cyclic protocols like the surface code

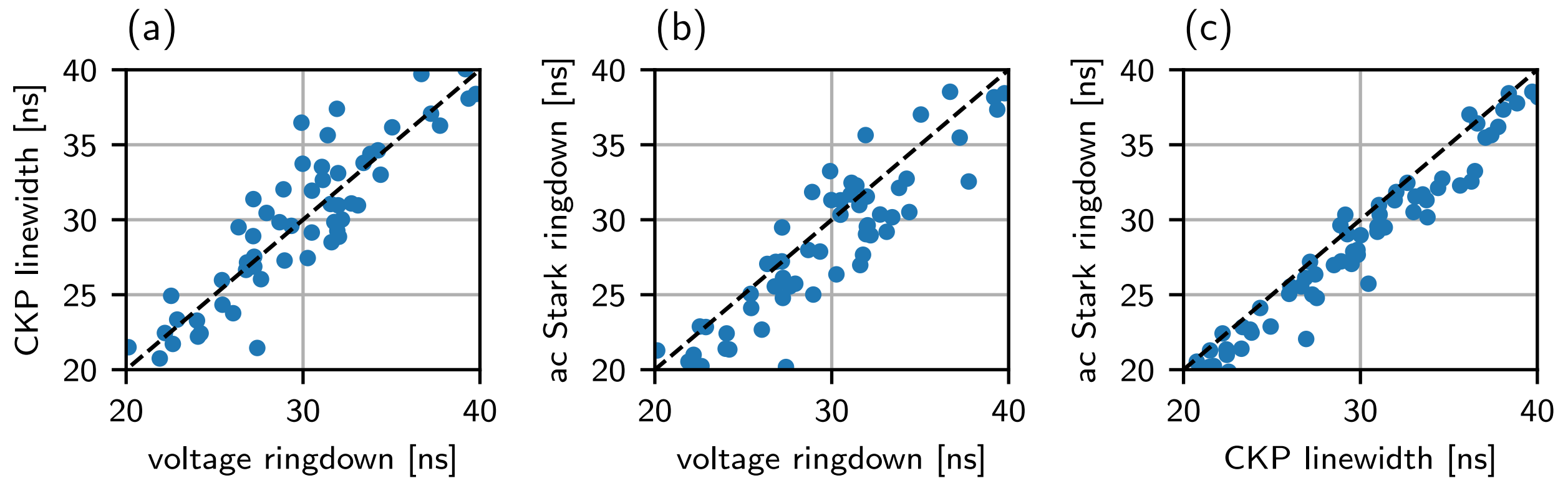

FIG. 4. Comparison of three methods of measuring $1/\kappa$. (a) Correlation between the CKP and voltage ringdown methods. (b) Correlation between the ac Stark ringdown and voltage ringdown methods. (c) Correlation between the ac Stark ringdown and CKP methods.

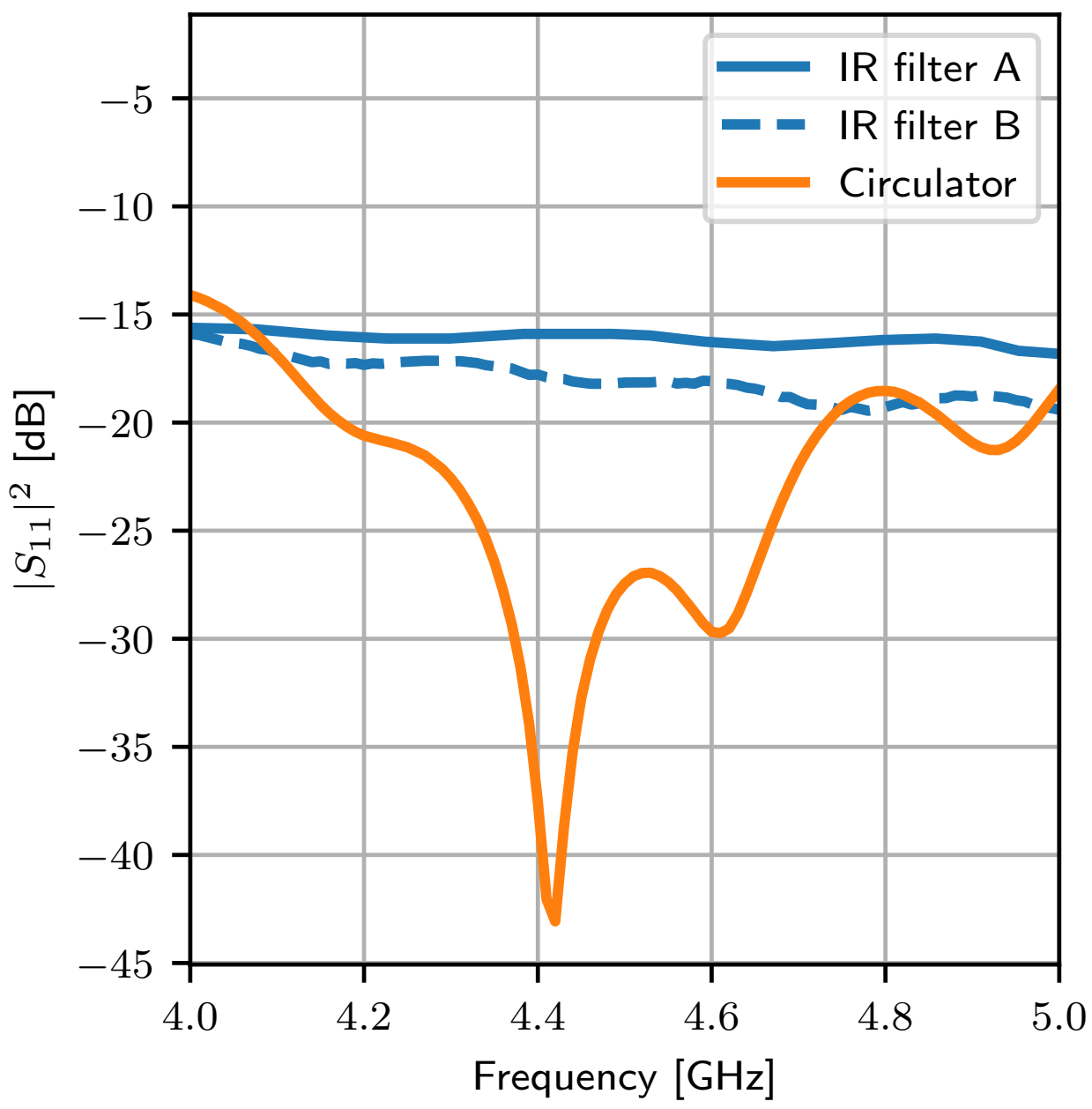

FIG. 5. Reflection coefficient of the IR filter and circulator.

[4], so even a few resonators in the low $\kappa$ tail of the distribution could be problematic.

We suspected that the root cause of the variance in $\kappa$ might be an impedance mismatch in the receiver, specifically coming from a reflection at the custom infrared (IR) filter or circulator closest to the qubit chip. We constructed a "cryo-vna" apparatus, similar to the one described in Ref. [23], enabling *in-situ* microwave calibration near the plane of the device under test and used it to measure the reflection coefficients of the IR filter and circulator. Results are shown in Fig. 5. If the receiver impedance mismatch were the only source of variation in $\kappa$, then the ratio of the largest and smallest observed values of $\kappa$ would be related to the receiver's reflection coefficient via the voltage standing wave ratio (VSWR)

$$\frac{\kappa_{\mathrm{max}}}{\kappa_{\mathrm{min}}} = \left(\frac{1+|S_{11}|}{1-|S_{11}|}\right)^2 = \mathrm{VSWR}^2 \, . \tag{8}$$

Near the center of our readout band at 4.8 GHz the IR filter reflection coefficient is as high as −16 dB (corresponding to a wave impedance mismatch in the range $[36\,\Omega, 68\,\Omega]$ in a 50Ω system) which, via Eq. (8), predicts $\kappa_{\mathrm{max}}/\kappa_{\mathrm{min}} \approx 1.9$, close to the observed value of 2. However, notice that the $|S_{11}|^2$ curves of the two IR filters shown in Fig. 5 differ by 4 dB in the readout band near 4.8 GHz, and note that the particular IR filters and circulators characterized in Fig. 5 are not the same physical units used in the experiments with qubits represented in Fig. 4. Therefore, we do not expect perfect agreement between the observed $\kappa$ distribution and the distribution predicted by Eq. (8) and values of $|S_{11}|$ shown in Fig. 5. Indeed, in another setup, we found that removing the IR filter and using a circulator with better impedance match reduced the spread in $\kappa$ from 3× to 2× (data not shown).

These findings suggest that impedance matching is a critical aspect of a multiplexed superconducting qubit readout system, as it plays a role in determining the system's maximum speed. In these systems, signal losses must be minimized, particularly between the qubit chip and the first amplifier (usually a Josephson parametric amplifier), so using attenuators to improve impedance match before the first amplifier is unlikely to be viable. These results highlight the importance of the microwave design of the sample package and readout receiver, which should be the subject of further study and development.

## V. READOUT EFFICIENCY

With $\chi$, $\kappa$, and the resonator drive power measured in the CKP experiment, we turn to the readout quantum efficiency $\eta$. The efficiency is the fraction of available information from the readout which is actually collected and contributes to the SNR. A common method for measuring quantum efficiency uses the fact that measuring the qubit energy necessarily causes the qubit to dephase [24, 25]. The magnitude of the off-diagonal element of the qubit density matrix $\rho_{10}$ is bounded by $|\rho_{10}| \leq (1/2)\exp(-\text{SNR}/4)$. For a given resonator drive pulse and measured $|\rho_{10}|$, the efficiency is the ratio of the measured SNR to the maximum SNR allowed by the inequality. This method has a few limitations. First, the coherence $\rho_{10}$ typically decays over time due to frequency noise sources independent of the readout signal, so the measured value of $|\rho_{10}|$ does not directly indicate the dephasing due to the resonator drive. Second, by construction, the SNR levels of interest for projective measurement correspond to tiny values of $|\rho_{10}|$ because projective measurement means that the qubit state is projected to one of the poles of the Bloch sphere where the phase coherence is zero. Therefore, it is awkward to use this strategy for measuring efficiency with the same resonator drive power levels as would actually be used in e.g. a quantum error correction code based on projective measurements.

Our approach is based on the fact that with $\chi$, $\kappa$ and power calibrated from the CKP experiment, we can predict the rate of increase of SNR with increasing resonator drive pulse length. With the rotating frame frequency at $\omega_{\text{r,m}}$ so that $\Delta = \mp\chi$, the steady state resonator fields for the qubit in states $|0\rangle$ and $|1\rangle$ are

$$\alpha_{|0\rangle,|1\rangle} = \frac{2A}{\sqrt{\kappa}} \frac{1}{1 \pm ix} \tag{9}$$

where $x = -2\chi/\kappa > 0$ and the upper and lower signs are for $|0\rangle$ and $|1\rangle$. The distance between the fields is

$$\left|\alpha_{|0\rangle} - \alpha_{|1\rangle}\right| = 4\frac{|A|}{\sqrt{\kappa}} \frac{x}{1 + x^2} \tag{10}$$

and the mean photon number in either state is

$$\bar{n} = |\alpha|^2 = \frac{4\,|A|^2}{\kappa} \frac{1}{1 + x^2} = \bar{n}^* \frac{1}{1 + x^2} \tag{11}$$

where $\bar{n}^* = 4\,|A|^2/\kappa$ is the number of photons that would be in the resonator if the resonator were driven on resonance. Combining Equations (4), (10), and (11), we find the rate of increase in SNR per time when the resonator is in the steady state

$$\frac{d\text{SNR}}{dt} = \eta\,\bar{n}\,\kappa \left(\frac{1}{1/2}\right) \left(\frac{4x^2}{1 + x^2}\right) \tag{12}$$

$$= \eta\,\bar{n}^*\,\kappa \left(\frac{1}{1/2}\right) \left(\frac{2x}{1 + x^2}\right)^2 . \tag{13}$$

The three factors in Eq. (12) can be roughly interpreted as the leaked photon flux, the noise, and the information per photon. With $\chi$, $\kappa$, and $\bar{n}$ known, a measurement of the "SNR flux" $d\text{SNR}/dt$ leaves $\eta$ as the only free parameter, which we find by fitting a line to the SNR flux versus time. To measure $d\text{SNR}/dt$, we prepare the qubit in either $|0\rangle$ or $|1\rangle$ and, drive the resonator for several times $1/\kappa$ to put it into the steady state, and then continue driving the resonator while collecting the dispersed signal for a variable time.

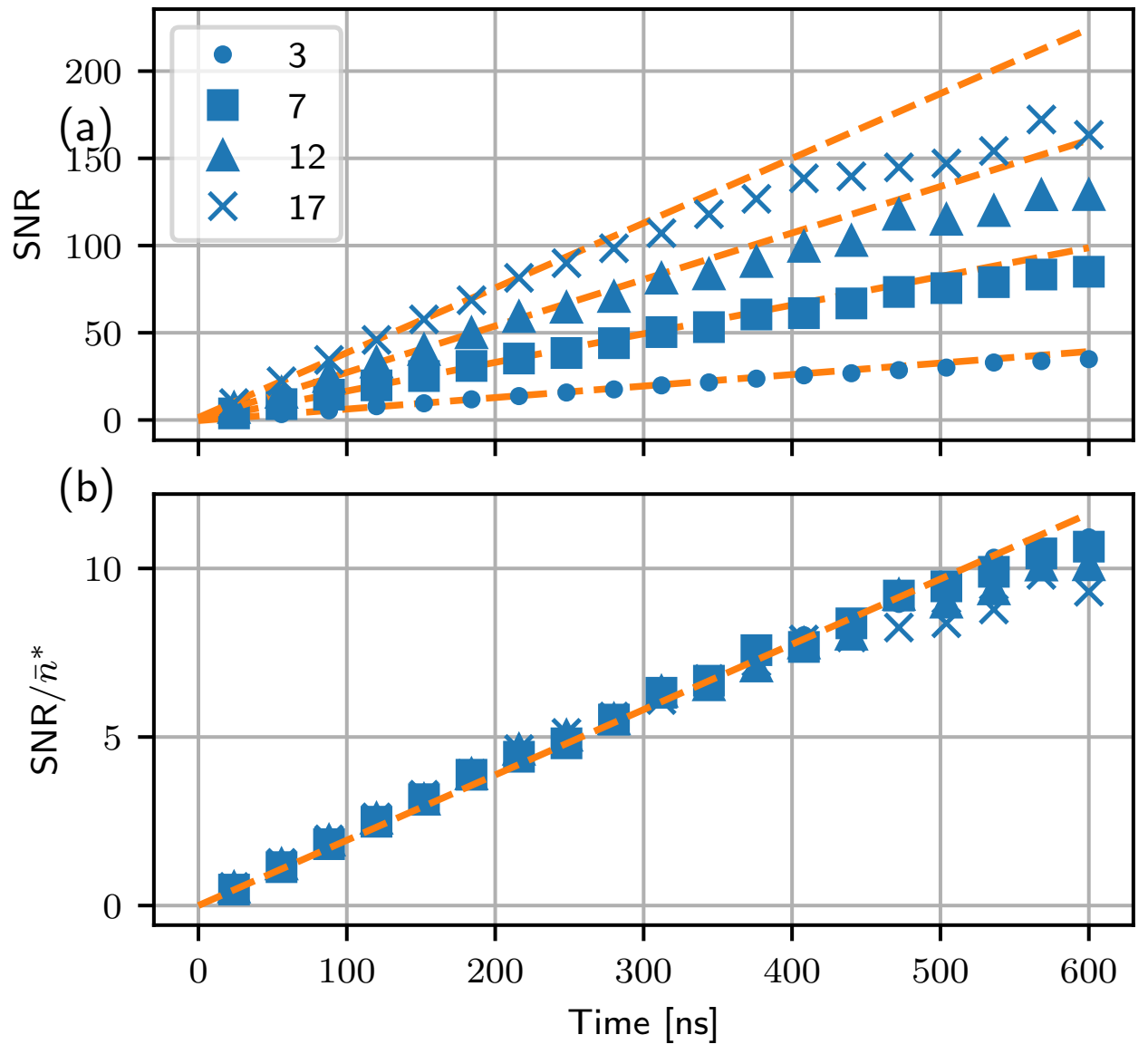

FIG. 6. (a) SNR versus integration time for several values of $\bar{n}^*$. The dashed lines are linear fits to the first 200 ns of data. (b) Same data but normalized by $\bar{n}^*$. The dashed line is a guide to highlight the curve bending.

Results are shown in Fig. 6. We repeat the experiment for a few different values of the drive power, observing that the SNR flux depends linearly on the drive power $\bar{n}^*$ as expected, i.e. we are not saturating the parametric amplifier. This approach was used in Ref. [26] to characterize the readout efficiency with SNIMPA parametric amplifiers. Note that with phase preserving amplifiers as used here, the maximum attainable efficiency is $\eta = 0.5$.

The data in Fig. 6 do not fall on completely straight lines. In some experiments, this happens because the qubit energy decay time ($T_1$) is too short, causing the cloud of points for $|1\rangle$ to distort toward the cloud for $|0\rangle$, reducing the extracted SNR. Interestingly, in some experiments we notice the curves bend over at a particular value of SNR regardless of drive power. The origin of this bending at a certain value of SNR is not understood.

We also study the statistical distribution of the measured efficiencies. Fig. 7 shows the cumulative distribution function of $\eta$ along with scatter plots of $\eta$ versus $1/\kappa$ and versus the parametric amplifier gain (gain is measured by observing the signal level as we turn the paramp pump on and off). Evidently, $\eta$ is uncorrelated with $1/\kappa$ as expected, having a Pearson correlation of 5%. On the other hand, we see increasing $\eta$ with increasing gain (in dB) with a Pearson correlation of 58%. This correlation of efficiency with gain is unsurprising; the receiver hardware (package, cabling, circulators) used here are similar to that used in Ref. [26]. Using equation (3) in that reference with $\alpha = 0.5$, $T_Q = T_P = 112\,\mathrm{mK}$, and $T_H = 2.5\,\mathrm{K}$, we estimate that as the paramp gain ranges from 10 dB to 20 dB, the efficiency should range approximately from 12% to 23%, in line with our observations here.

## VI. MODEL AND MEASUREMENTS OF READOUT SNR

With our system parameters measured, we come to comparison of predicted and measured SNR. We measure the SNR in single qubit readout on 54 qubits in a Sycamore-style device. Whereas the system characterization experiments discussed above are done with the resonator in the steady state, here we use readout pulses. For each qubit, we measure the system parameters described above with the qubit at two different frequencies, which we refer to as reference points "A" and B". We then vary the qubits' frequencies in a 360 MHz band around these reference points. At each frequency we tune up readout, using Eq. (4) and the system parameters from the reference point, targeting SNR = 15 (corresponding to a separation error [6] of 0.3%). We repeat this procedure for three different pulse durations. For each qubit frequency and pulse duration, we model the expected SNR via Eq. (4). We plot the disagreement between the measured and modelled SNR, marginalized over the qubit frequencies, in Figure 8, finding most of the data within ±10%. Interestingly, the prediction seems to be the least accurate for the shortest pulse duration (200 ns) where the measured SNR is somewhat *larger* than predicted by the model. In any case, the agreement between experiment and model is good enough that we continue to use the model as the basis for development of more advanced designs.

## VII. CONCLUSIONS

We have shown methods for measuring the four most important parameters in a dispersive readout system: the disersive shift $\chi$, the resonator linewidth $\kappa$, the drive power $A$, and the quantum efficiency $\eta$. The dispersive shift, resonator line width, and drive power were measured in a single self-calibrating procedure (CKP). Three different measurements of $\kappa$ – voltage ringdown, ac Stark ringdown, and CKP – gave similar results with CKP and ac Stark ringdown agreeing the best. The efficiency can was measured by comparing predicted and observed SNR. Combining the measured parameters into a simple model predicted the SNR in single qubit readout on a Sycamore processor with 54 qubits to ±10%. We found large variation in $\kappa$ and that attention to impedance matching in the readout receiver reduces this variation. Further improvements to the variation in $\kappa$ may be required for future fault tolerant quantum computing systems.

Two future directions suggested by this work are 1) Improved control over $\kappa$ and 2) modelling and measurement of multiplexed readout with sufficient speed for fault tolerant quantum computing, in a device also supporting single- and two-qubit gates sufficiently accurate for fault tolerant quantum computation.

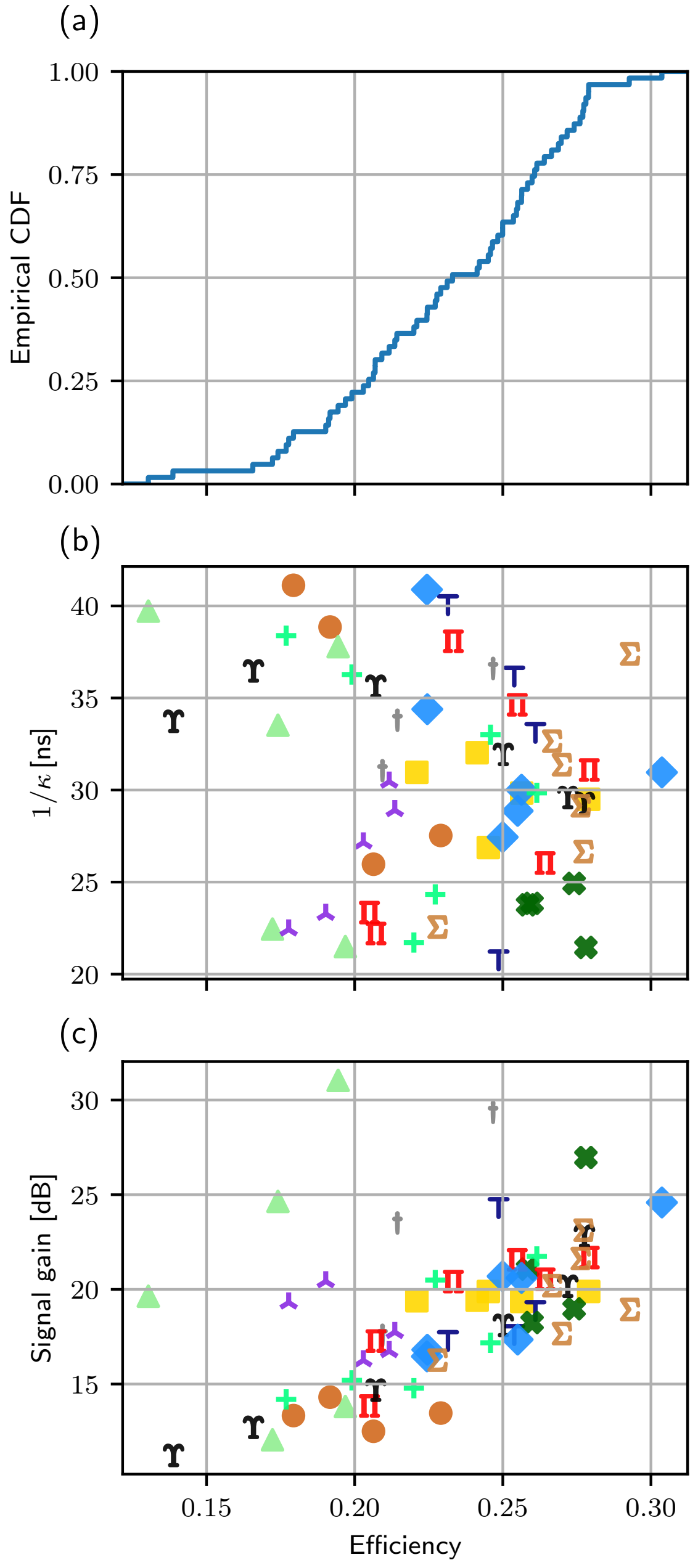

FIG. 7. (a) Cumulative distribution function of the readout efficiency. (b) Readout efficiency versus $1/\kappa$. (c) Readout efficiency versus parametric amplifier gain. In (b) and (c) each marker symbol corresponds to a different multiplexed readout line on the chip.

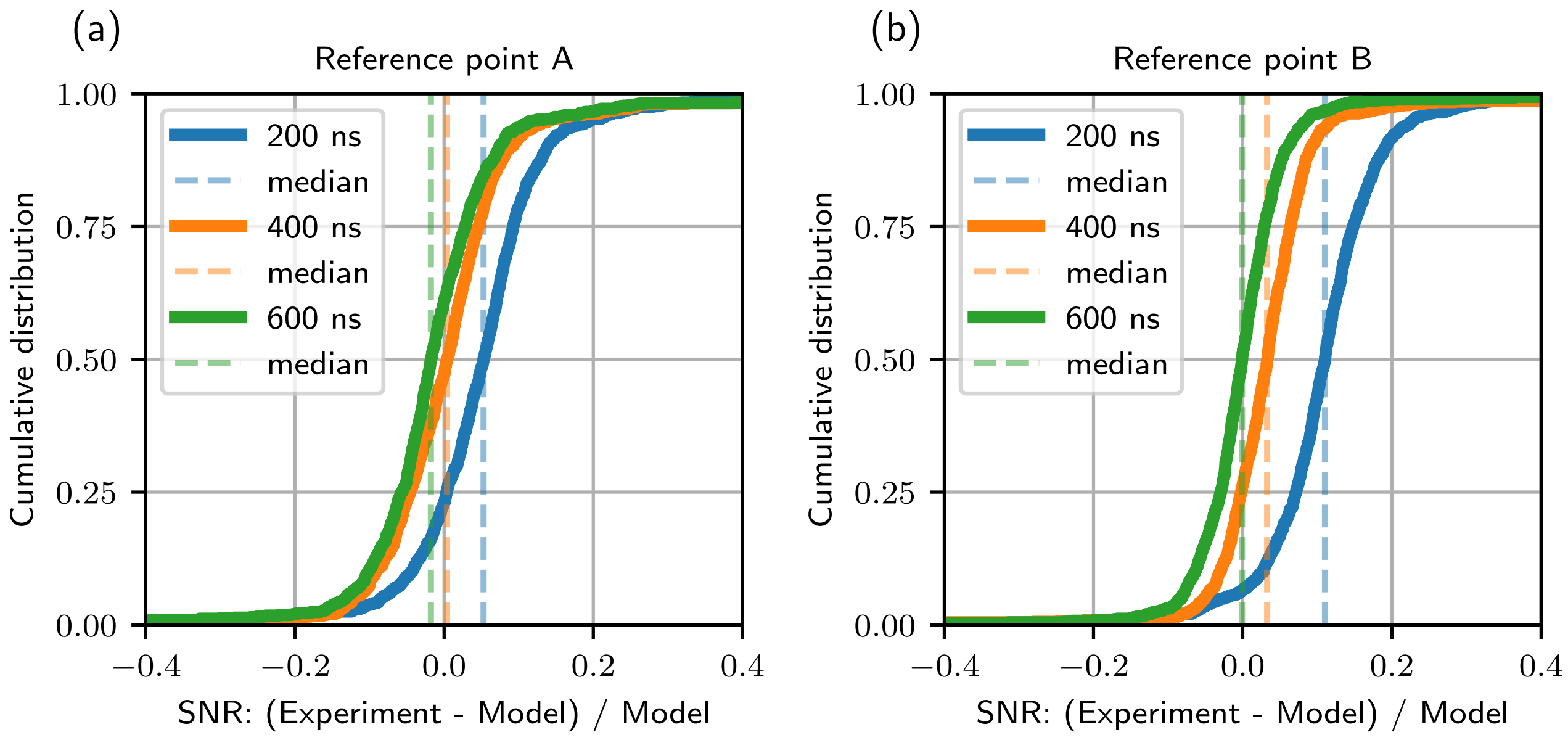

FIG. 8. Cumulative distribution of the relative error in measured versus predicted readout SNR. (a) Qubits calibrated at reference point A. (b) Qubits calibrated at reference point B.

## VIII. ACKNOWLEDGEMENTS

We thank Ebrahim Forati, Reza Fatemi, Vlad Sivak, Clarke Smith, William Livingston, and Juan Atalaya for critical readings of the manuscript. The Google Quantum AI team fabricated the processor, built the cryogenic and control systems, optimized the processor performance, and provided the tools that enabled execution of this experiment.

## Appendix A: SNR model

This appendix spells out the details of the readout receiver, including analog mixing and filtering, and digital signal processing. The discussion centers around derivation of Equation (4) via frequency domain analysis with attention drawn to the various assumptions and limits under which that equation is correct. While the analysis includes consideration of discrete time sampling, we will see that the analytic result is valid precisely when the filtering and sampling are designed to preserve the properties of the incoming analog signals and noise. All integrals in this appendix are taken over the interval $[-\infty, \infty]$ unless noted otherwise.

A schematic of the receiver is shown in Fig. 9. The analog signal (and noise) come in on a transmission line with characteristic impedance $Z_0$. The signal is mixed to intermediate frequencies in an IQ mixer, filtered in lowpass anti-aliasing filters with frequency response $\tilde{h}$, sampled at discrete times spaced by $\delta t$, and then digitally processed to produce a single point $Z$ in the complex "IQ" plane.

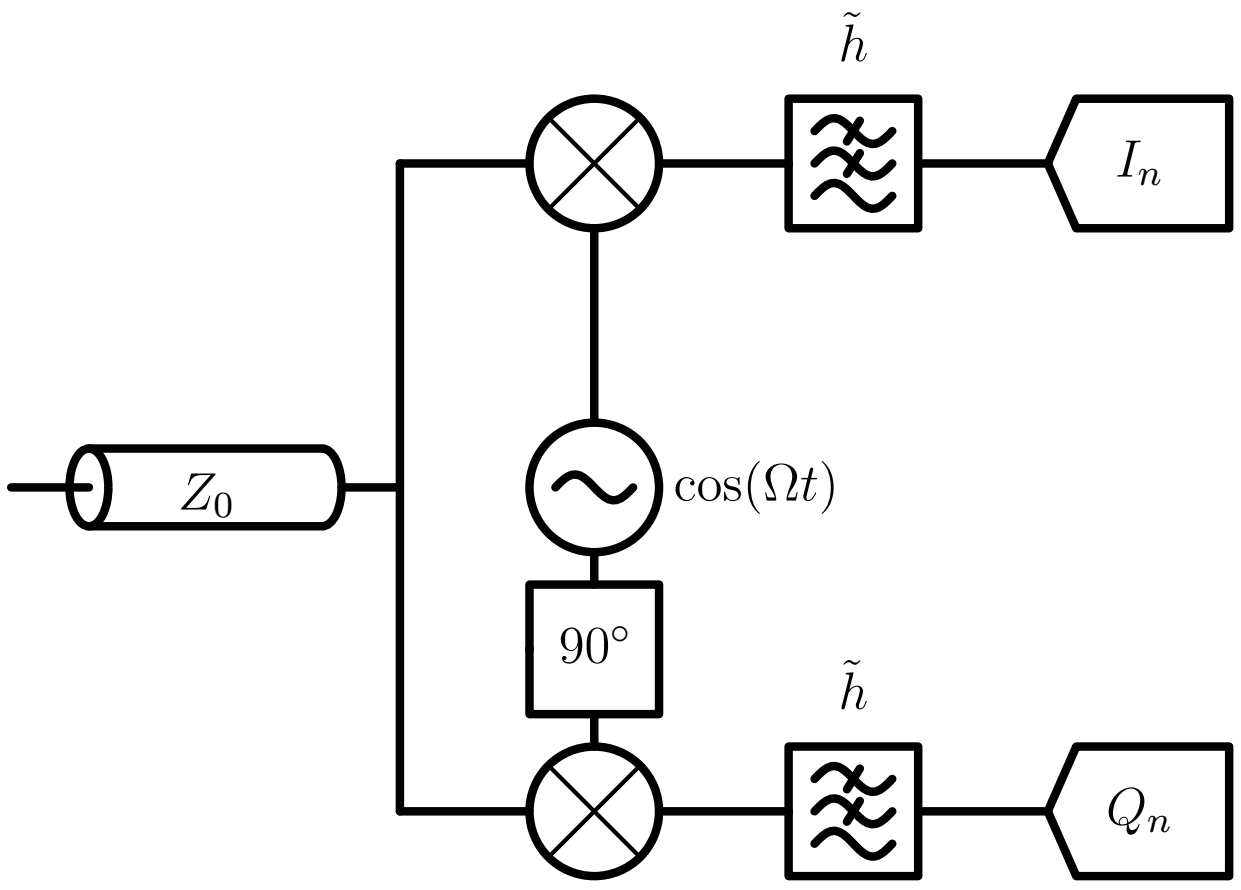

FIG. 9. The readout receiver. The signal and noise enter from the left along a transmission line. They are mixed to the intermediate frequency in an IQ mixer, here modelled as a local oscillator driving two mixers with a 90° phase shift. The intermediate frequency signals are lowpass flitered and then digitized.

### 1. Signal

Let the incoming microwave voltage pulse be denoted $V_{\mu\text{w}}$. Because $V_{\mu\text{w}}$ is real, it can be written as a Fourier transform over positive frequencies

$$\begin{aligned} V_{\mu\text{w}}(t) = &\int_0^\infty \frac{d\omega}{2\pi}\tilde{V}_{\mu\text{w}}(\omega)e^{i\omega t} \\ &+ \int_0^\infty \frac{d\omega}{2\pi}\tilde{V}_{\mu\text{w}}(\omega)^* e^{-i\omega t}\,. \end{aligned} \tag{A1}$$

If $V_{\mu\text{w}}$ has nonzero content only in a band $[\Omega - B, \Omega + B]$ then its Fourier integral representation can be rewritten by a change of variables $\omega = \Omega + \delta\omega$ as ($\delta\omega$ is the detuning away from the local oscillator)

$$\begin{aligned} V_{\mu\text{w}}(t) = &\int_{-B}^{B} \frac{d\delta\omega}{2\pi}\big(\tilde{V}_{\mu\text{w}}(\Omega+\delta\omega)e^{i(\Omega+\delta\omega)t} \\ &+\tilde{V}_{\mu\text{w}}(\Omega+\delta\omega)^* e^{-i(\Omega+\delta\omega)t}\big) \qquad \text{(A2)} \\ = &\,2\text{Re}\Big[\int_{-B}^{B}\frac{d\delta\omega}{2\pi}\tilde{V}_{\mu\text{w}}(\Omega+\delta\omega)e^{i(\Omega+\delta\omega)t}\Big]\,. \qquad \text{(A3)} \end{aligned}$$

The microwave signal is passed through the IQ mixer, resulting in two intermediate frequency (IF) signals

$$\begin{aligned} I_{\text{mixer}}(t) &= A\cos(\Omega t)V_{\mu\text{w}}(t) \\ &= A\,\text{Re}\left[\int_{-B}^{B}\frac{d\delta\omega}{2\pi}\tilde{V}_{\mu\text{w}}(\Omega+\delta\omega)e^{i\delta\omega t}\right] \qquad \text{(A4)} \\ Q_{\text{mixer}}(t) &= -A\sin(\Omega t)V_{\mu\text{w}}(t) \\ &= A\,\text{Im}\left[\int_{-B}^{B}\frac{d\delta\omega}{2\pi}\tilde{V}_{\mu\text{w}}(\Omega+\delta\omega)e^{i\delta\omega t}\right] \qquad \text{(A5)} \end{aligned}$$

where here the factor $A$ accounts for conversion loss in the mixer and we have dropped the higher frequency ($2\Omega$) mixing products because they are filtered away in the next step. The IF signals are passed through low pass (anti-aliasing) filters resulting in

$$\begin{aligned} I(t) &= \\ &A\,\text{Re}\left[\int_{-B}^{B}\frac{d\delta\omega}{2\pi}\tilde{V}_{\mu\text{w}}(\Omega+\delta\omega)\tilde{h}(\delta\omega)e^{i\delta\omega t}\right] \qquad \text{(A6)} \\ Q(t) &= \\ &A\,\text{Im}\left[\int_{-B}^{B}\frac{d\delta\omega}{2\pi}\tilde{V}_{\mu\text{w}}(\Omega+\delta\omega)\tilde{h}(\delta\omega)e^{i\delta\omega t}\right]\,. \qquad \text{(A7)} \end{aligned}$$

Note that $\tilde{h}$ is dimensionless. Because $I(t)$ and $Q(t)$ are the real and imaginary parts of the same quantity, it is mechanically and conceptually convenient to define $Z(t) = I(t) + iQ(t)$, or in the frequency domain $\tilde{Z}(\omega) = A\tilde{V}_{\mu\text{w}}(\Omega+\omega)\tilde{h}(\omega)$.

Given a frame rotating with frequency $\Omega$, we define the rotating frame representation $\bar{V}_{\mu\text{w}}$ of the microwave signal by the equation $V_{\mu\text{w}}(t) = \text{Re}\left[e^{i\Omega t}\bar{V}_{\mu\text{w}}(t)\right]$, from which we find via Eq. (A3)

$$\bar{V}_{\mu\text{w}}(t) = 2\int_{-B}^{B}\frac{d\delta\omega}{2\pi}\tilde{V}_{\mu\text{w}}(\Omega+\delta\omega)e^{i\delta\omega t} \tag{A8}$$

or $\tilde{\bar{V}}_{\mu\text{w}}(\omega) = 2\tilde{V}_{\mu\text{w}}(\Omega+\omega)$. The post-filter IF signal can then be written as

$$\tilde{Z}(\omega) = \frac{A}{2}\tilde{\bar{V}}_{\mu\text{w}}(\omega)\tilde{h}(\omega)\,. \tag{A9}$$

The IF signals is digitally sampled forming the sequence

$$Z_n = D\int\frac{d\omega}{2\pi}\tilde{Z}(\omega)e^{i\omega n\delta t} \tag{A10}$$

where $D$ is a conversion factor from voltage to digital amplitude ($D$ has dimensions of 1/Voltage) and $\delta t$ is the sampling interval. In the work described here $\delta t = 1\,\text{ns}$. The demodulated IQ point $Z$ is computed digitally as

$$\begin{aligned} Z &= \frac{1}{N}\sum_{n=0}^{N-1} Z_n e^{-i\,\omega_{\text{sb}}\,n\delta t}W_n \\ &= \frac{AD}{2N}\sum_{n=0}^{N-1}\int\frac{d\omega}{2\pi}\tilde{\bar{V}}_{\mu\text{w}}(\omega)\tilde{h}(\omega)e^{i\omega n\delta t}e^{-i\omega_{\text{sb}}\,n\delta t}W_n \end{aligned} \tag{A11}$$

where $\omega_{\text{sb}}$ is the sideband frequency of the channel and $W_n$ defines a "window function" used to optimize the SNR [15, 16]. We define a dimensionless variable $x = (\omega - \omega_{\text{sb}})\delta t$ and move the sum inside the integrals to get

$$\begin{aligned} Z = &\frac{AD}{2N\delta t}\int\frac{dx}{2\pi}\tilde{\bar{V}}_{\mu\text{w}}(x/\delta t + \omega_{\text{sb}})\tilde{h}(x/\delta t + \omega_{\text{sb}}) \\ &\sum_{n=0}^{N-1} e^{inx}W_n\,. \end{aligned} \tag{A12}$$

We now assume that $W_n$ can be interpolated as a smooth function $W : \mathbb{R} \to \mathbb{R}$. The discrete samples $W_n$ can then be written in the frequency domain

$$W_n = \frac{1}{\delta t}\int\frac{dy}{2\pi}\tilde{W}(y/\delta t)e^{iyn}\,. \tag{A13}$$

Putting everything together, we have

$$\begin{aligned} Z = &\frac{AD}{2N\delta t^2}\int\frac{dx}{2\pi}\tilde{\bar{V}}_{\mu\text{w}}(x/\delta t + \omega_{\text{sb}})\tilde{h}(x/\delta t + \omega_{\text{sb}}) \\ &\int\frac{dy}{2\pi}\tilde{W}(y/\delta t)\sum_{n=0}^{N-1} e^{i(x+y)n}\,. \end{aligned} \tag{A14}$$

The sum is the known Dirichlet kernel $K$ defined as $K(N, u) = \sin(Nu/2)/\sin(u/2)$. In the limit $N \to \infty$ the Dirichlet kernel approaches a comb of Dirac delta functions $K(N \to \infty, u) = \sum_{k=-\infty}^{\infty}(2\pi)\delta(u - 2\pi k)$. Our first assumption is that $N$ is large enough to use this

Dirac comb approximation. We can now do the integral over $y$ to find

$$Z = \frac{AD}{2N\delta t^2}\int \frac{dx}{2\pi}\sum_{k=-\infty}^{\infty} \tilde{\bar{V}}_{\mu\mathrm{w}}(x/\delta t + \omega_{\mathrm{sb}}) \tilde{h}(x/\delta t + \omega_{\mathrm{sb}})\tilde{W}((2\pi k - x)/\delta t) \tag{A15}$$

and having handled the sum over $n$, we switch back to $\omega = x/\delta t$

$$Z = \frac{AD}{2N\delta t}\int \frac{d\omega}{2\pi}\sum_{k=-\infty}^{\infty} \tilde{\bar{V}}_{\mu\mathrm{w}}(\omega + \omega_{\mathrm{sb}}) \tilde{h}(\omega + \omega_{\mathrm{sb}})\tilde{W}(2\pi k/\delta t - \omega)\,. \tag{A16}$$

In a frequency multiplexed system, the microwave signal in the rotating frame has several components

$$\bar{V}_{\mu\mathrm{w}}(t) = \sum_{\text{channel } c} e^{i\omega_{\mathrm{sb},c}t}\bar{V}^{c}_{\mu\mathrm{w}}(t) \tag{A17}$$

where each $\bar{V}^{c}_{\mu\mathrm{w}}$ is a slowly varying envelope and each component in the sum is meant to probe one qubit's readout resonator. In the frequency domain,

$$\tilde{\bar{V}}_{\mu\mathrm{w}}(\omega) = \sum_{\text{channel } c} \tilde{\bar{V}}^{c}_{\mu\mathrm{w}}(\omega - \omega_{\mathrm{sb},c})\,. \tag{A18}$$

In this case, we find

$$Z = \frac{AD}{2N\delta t}\int \frac{d\omega}{2\pi}\sum_{k=-\infty}^{\infty}\sum_{\text{channel } c} \tilde{\bar{V}}^{c}_{\mu\mathrm{w}}(\omega + \omega_{\mathrm{sb}} - \omega_{\mathrm{sb},c}) \tilde{h}(\omega + \omega_{\mathrm{sb}})\tilde{W}(2\pi k/\delta t - \omega)\,. \tag{A19}$$

Now we make five further assumptions. First, we assume that the various $\tilde{\bar{V}}^{c}_{\mu\mathrm{w}}$ do not appreciably overlap. The degree to which this assumption is valid is a critical aspect of readout design. As multiplexing factors are increased and readout pulse duration is decreased, keeping the various frequency channels from overlapping requires separating them further in frequency space, which requires higher bandwidth amplifiers, sharp cutoff Purcell filters, larger qubit-resonator detuning, and consequently larger qubit-resonator coupling. While we do not study this issue in detail here, we mention that we have studied this issue analytically, numerically, and experimentally, finding that for realistically achievable parameters, multiplexing beyond 10 qubits with readout fast enough for error correction will be a challenge. In short, frequency crowding is likely to be one of the most important challenges in superconducting qubit readout, and should be investigated in depth. Second, we assume that the total bandwidth occupied by all $\tilde{\bar{V}}^{c}_{\mu\mathrm{w}}$ is smaller than $(2\pi)/\delta t$. Third, we choose $\omega_{\mathrm{sb}}$ to be equal to one of the $\omega_{\mathrm{sb},c}$, which we denote $\omega_{\mathrm{sb},c^*}$. This is just saying that the sideband frequency used in digital mixing is chosen to center one of the frequency channels at dc. Fourth, we assume that $W$ is chosen to "window" just that centered frequency channel. By construction $\tilde{h}(x/\delta t)$ is chosen to be nonzero only for $|x| \lesssim \pi/2$ [27]. Under these conditions, the only nonzero term in the integrand is for $k = 0$. See Fig. 10. Fifth and finally, we assume that $\tilde{h}$ varies sufficiently slowly over the width of $\tilde{W}$ such that we can replace $\tilde{h}(\omega + \omega_{\mathrm{sb}})$ with $\tilde{h}(\omega_{\mathrm{sb},c^*})$. With these assumptions in place we have

$$\begin{aligned} Z &= \frac{AD}{2N\delta t}\tilde{h}(\omega_{\mathrm{sb},c^*})\int \frac{d\omega}{2\pi}\tilde{\bar{V}}^{c^*}_{\mu\mathrm{w}}(\omega)\tilde{W}(-\omega) \\ &= \frac{AD}{2N\delta t}\tilde{h}(\omega_{\mathrm{sb},c^*})\int dt\, \bar{V}_{\mu\mathrm{w}}(t)W(t)\,. \end{aligned} \tag{A20}$$

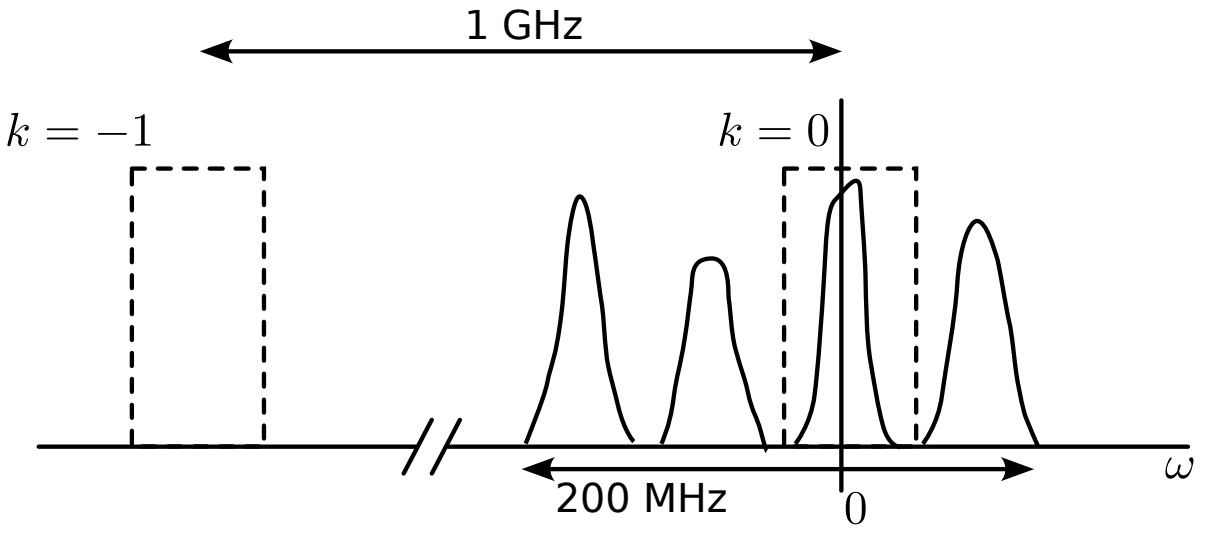

FIG. 10. Drawing of the frequency domain integrand in Equation (A19). The peaks drawn in solid line represent the frequency components $\tilde{\bar{V}}^{c}_{\mu\mathrm{w}}$ and the dashed line boxes represent $\tilde{W}(2\pi k/\delta t - \omega)$ for $k = 0$ and $-1$.

To make connection with the main text, note that the power flowing into the receiver in terms of the voltage is $\bar{V}^2_{\mu\mathrm{w}}/2Z_0$, while in terms of the amplitude wave $B(t)$ from the main text the power is $\hbar\omega\,|B|^2$. Therefore, $\bar{V}_{\mu\mathrm{w}}(t) = \sqrt{2Z_0\hbar\omega}B(t)$ and

$$\begin{aligned} \left|Z_{|0\rangle} - Z_{|1\rangle}\right| &= \frac{AD}{2N\delta t}\sqrt{2Z_0\hbar\omega} \\ &\quad \left|\int dt\, \left(B_{|0\rangle}(t) - B_{|1\rangle}(t)\right) W(t)\right|\,. \end{aligned} \tag{A21}$$

Evidently, the constant $C$ from the main text is $C = (AD/2N\delta t)\sqrt{2Z_0\hbar\omega}$.

### 2. Noise

To complete our formula for the SNR, we need to find the width of the cloud of points in the IQ plane. Our approach is to write out the IF wave forms in the case where the incoming microwaves is noise with a known spectral density. We then explicitly compute the average of the square modulus of the demodulated IQ point. The purpose of this approach is to highlight assumptions and associated approximations.

Suppose that along with our microwave voltage signal comes voltage noise denoted $V_\xi$. The IF noise wave forms

after the filter are

$$I(t) = A\int dt'\, V_\xi(t')\cos(\Omega t')h(t-t') \tag{A22}$$

$$Q(t) = -A\int dt'\, V_\xi(t')\sin(\Omega t')h(t-t') \tag{A23}$$

and the digital complex signal is

$$Z_n = AD\int dt'\, V_\xi(t')e^{-i\Omega t'}h(n\delta t-t')\,. \tag{A24}$$

The demodulated complex point $Z$ is a weighted sum of random variables

$$\begin{aligned} Z =& \frac{1}{N}\sum_{n=0}^{N-1} Z_n e^{-i\omega_{\mathrm{sb}} n\delta t}W_n^c \\ =& \frac{AD}{N}\sum_{n=0}^{N-1}\int dt' \\ & V_\xi(t')e^{-i\Omega t'}e^{-i\omega_{\mathrm{sb}} n\delta t}h(n\delta t-t')W_n^c \\ =& \frac{AD}{N}\sum_{n=0}^{N-1}\int dt' \\ & V_\xi(n\delta t-t')e^{-i\Omega(n\delta t-t')}e^{-i\omega_{\mathrm{sb}} n\delta t}h(t')W_n^c\,. \end{aligned} \tag{A25}$$

The mod square of $Z$ is

$$\begin{aligned} |Z|^2 = & \left(\frac{AD}{N}\right)^2 \sum_{n,m=0}^{N-1}\int dt'\,dt'' \\ & V_\xi(n\delta t-t')V_\xi(m\delta t-t'')h(t')h(t'') \\ & e^{i\Omega(t'-t'')}e^{-i(\Omega+\omega_{\mathrm{sb}})(n-m)\delta t}W_n^c W_m^{c\,*}\,. \end{aligned} \tag{A26}$$

Now taking the statistical average we have

$$\begin{aligned} \left\langle |Z|^2\right\rangle = & \left(\frac{AD}{N}\right)^2 \sum_{n,m=0}^{N-1}\int dt'\,dt'' \\ & \langle V_\xi(n\delta t-t')V_\xi(m\delta t-t'')\rangle\, h(t')h(t'') \\ & e^{i\Omega(t'-t'')}e^{-i(\Omega+\omega_{\mathrm{sb}})(n-m)\delta t}W_n^c W_m^{c\,*}\,. \end{aligned} \tag{A27}$$

The Weiner-Kinchien theorem relates the two-point average noise to the single-sided noise spectral density

$$\langle V_\xi(t)V_\xi(0)\rangle = \frac{1}{2}\int\frac{d\omega}{2\pi}S_{V_\xi}(\omega)e^{i\omega t}\,. \tag{A28}$$

Plugging into our formula for $\left\langle |Z|^2\right\rangle$ we have

$$\begin{aligned} \left\langle |Z|^2\right\rangle =& \frac{1}{2}\left(\frac{AD}{N}\right)^2 \sum_{n,m=0}^{N-1}\int dt'\,dt''\,\frac{d\omega}{2\pi} \\ & S_{V_\xi}(\omega)e^{i\omega((n-m)\delta t-(t'-t''))}h(t')h(t'') \\ & e^{i\Omega(t'-t'')}e^{-i(\Omega+\omega_{\mathrm{sb}})(n-m)\delta t}W_n^c W_m^{c\,*}\,. \end{aligned} \tag{A29}$$

The integrals over $t$ and $t'$ can be evaluated as Fourier transforms of the filter $h$, leaving

$$\begin{aligned} \left\langle |Z|^2\right\rangle =& \frac{1}{2}\left(\frac{AD}{N}\right)^2 \sum_{n,m=0}^{N-1}\int\frac{d\omega}{2\pi} \\ & S_{V_\xi}(\omega)e^{i\omega(n-m)\delta t}\left|\tilde{h}(\omega-\Omega)\right|^2 \\ & e^{-i(\Omega+\omega_{\mathrm{sb}})(n-m)\delta t}W_n^c W_m^{c\,*}\,. \end{aligned} \tag{A30}$$

Now we push the sum inside and define $x = \delta t(\omega-\Omega-\omega_{\mathrm{sb}})$ giving

$$\begin{aligned} \left\langle |Z|^2\right\rangle =& \frac{(AD)^2}{2N^2\delta t}\int\frac{dx}{2\pi}S_{V_\xi}\left(\Omega+\omega_{\mathrm{sb}}+x/\delta t\right) \\ & \left|\tilde{h}(x/\delta t+\omega_{\mathrm{sb}})\right|^2\left|\sum_{n=0}^{N-1}e^{inx}W_n\right|^2\,. \end{aligned} \tag{A31}$$

As we did for the signal, we assume that $W_n$ can be extrapolated by a continuous function, express it in the frequency domain, and approximate the sum (the Dirichlet kernel) as a comb of delta functions. With the same assumptions we made for the signal, we find

$$\begin{aligned} \left\langle |Z|^2\right\rangle =& \frac{(AD)^2}{2N^2\delta t^3}\int\frac{dx}{2\pi}S_{V_\xi}(\Omega+\omega_{\mathrm{sb}}+x/\delta t) \\ & \left|\tilde{h}(x/\delta t+\omega_{\mathrm{sb}})\right|^2\left|\tilde{W}(-x/\delta t)\right|^2\,. \end{aligned} \tag{A32}$$

$\tilde{W}(x)$ limits the integration over $x$ to a narrow range, so we approximate $\tilde{h}$ as a constant over that range, and similarly for $S_{V_\xi}$, leaving

$$\begin{aligned} \left\langle |Z|^2\right\rangle =& S_{V_\xi}(\Omega+\omega_{\mathrm{sb}})\left|\tilde{h}(\omega_{\mathrm{sb}})\right|^2\frac{(AD)^2}{2N^2\delta t^3} \\ & \int\frac{dx}{2\pi}\left|\tilde{W}(-x/\delta t)\right|^2 \\ =& S_{V_\xi}(\Omega+\omega_{\mathrm{sb}})\left|\tilde{h}(\omega_{\mathrm{sb}})\right|^2\frac{(AD)^2}{2(N\delta t)^2} \\ & \int dt\,|W(t)|^2\,. \end{aligned} \tag{A33}$$

The quantum limit of voltage noise spectral density at frequency $\omega$ is $S_{V_\xi} = Z_0\hbar\omega/2$. Note also that while we computed $\left\langle |Z|^2\right\rangle$, the width $\sigma$ in Eq. (4) is the width of a 1-dimensional projection (marginal distribution) of the 2-dimensional distribution of $Z$. If $Z$ is Gaussian, then $\sigma^2 = \left\langle |Z|^2\right\rangle/2$. Collecting these facts, we have

$$2\sigma^2 = \frac{1}{4}\frac{(AD)^2 Z_0\hbar\Omega}{(N\delta t)^2}\left|\tilde{h}(\omega_{\mathrm{sb}})\right|^2\int|W(t)|^2\,dt\,. \tag{A34}$$

### 3. Signal to noise ratio

Combining Equations (A21) and (A34), the signal to noise ratio is

$$\mathrm{SNR} = \frac{\left|Z_{|0\rangle} - Z_{|1\rangle}\right|^2}{2\sigma^2} = \kappa \frac{\left|\int dt\, \left(\alpha_{|0\rangle}(t) - \alpha_{|1\rangle}(t)\right) W(t)\right|^2}{(1/2)\int dt\, |W(t)|^2} \tag{A35}$$

where we used $B(t) = \sqrt{\kappa}\alpha(t)$. This result agrees with Eq. (4) up to the factor of $\eta$ which can be inserted by hand to represent either signal loss or excess noise above the quantum limit.